\documentstyle[twocolumn,aps]{revtex} 
\input psfig
\begin{document}
\title{ Study of the Phase Diagram and the $T_c$ Pressure Dependence in the 
$HgBa_2CuO_{4-\delta}$ Superconductors     }
\author{E. V.L. de Mello and C. Acha} 
\address{Centre National de la Recherche Scientifique,\\ Centre de Recherche sur les Tr\`{e}s Basses Temp\'{e}ratures\\
        Boite Postale 166, 38042 Grenoble-Cedex, France. \\}
\maketitle
\begin{abstract}
 We  apply a method recently derived  based on the extended Hubbard
Hamiltonian to obtain the superconducting critical temperature  
dependence with carrier density of $HgBa_2CuO_{4-\delta}$.
We show that this approach can be used to study the effects of pressure
and reproduces for the first time 
the experimental data for the whole
doping regime and their high and low pressure dependence.
\end{abstract}


One of the effects of the  pressure which is generally 
accepted is an increase of the carrier concentration on the $CuO_2$
planes transferred from the reservoir layers. Such pressure induced
charge transfer (PICT) has been confirmed by Hall effect and
thermoeletric power measurements in several compounds\cite{Taka}.
Therefore this effect was largely explored to account for the 
quantitative relation between $T_c$ and the pressure P and gave
origin to several models\cite{Almasan,Neumeier,Gupta}. Some of 
these models also invoked that several HTSs have a $T_{c}$ versus
carrier density  $n$ (per $CuO_2$) diagram which satisfies 
a phenomenological
universal parabolic curve, i.e., $T_{c}=T_c^{max}[1-\eta(n-n_{op})^{2}]$
where $n_{op}$ is the optimum $n$. Following along these lines we
can write $T_{c}=T_{c}(n,P)$ and easily derive a expansion in 
powers of P, namely 
\begin{equation}
 T_{c}(n,P)=T_{c}(n,0)+\alpha_1(n)P+\alpha_2(n)P^2
\label{exp1}
\end{equation}
where the coefficients are, 
\begin{equation}
\alpha_1(n)=\partial T_{c}(n)/\partial{P}
-2\eta T_c^{max}(n-n_{op})\partial{n}/\partial{P} 
\label{alpha1}
\end{equation}
in which the first term is known as the intrinsic term  and $\alpha_2(n)=
-\eta T_c^{max}(\partial{n}/\partial{P})^2$. This simple method
was very successful in describing the data in the vicinity of 
$n_{op}$\cite{Neumeier}
but fails to describe more recently set of data with a large variation
of $n$ values from underdoped to overdoped regime\cite{Taka,Cao}.
Futhermore this phenomenological approach does not give any indication
of the physical origin of the intrinsic term. 

We propose in this work some new ideas to interpret the effects of 
pressure. We apply a recently introduced approach\cite{Mello}
based on a BCS type mean field analysis  which uses the 
extended Hubbard Hamiltonian (t-U-V)
on a square lattice (of lattice parameter a).

In connection with this method we are led to propose that
the effects of pressure are two-fold: (i)- The well accepted PICT;
(ii)- The relation $2\Delta_{0}=\gamma K_{B}T_c^{max}$ ($\gamma=3.5$
for weakly BCS and $\gamma\approx4.3$ for the method
mentioned above)
suggests a very definite effect of the pressure on $\Delta_0$. This
is equivalent to say that the structural modifications due to the applied
pressure cause a variation on the attractive potential V (which is the
most influencial parameter on the value of $\Delta_0$). In fact the real effect
of the structural changes can only be estimate by electronic band
calculations\cite{Taka,Schill,Novikov},  but they are not  adequate to 
be performed in the presence of
the strong correlations of the t-U-V Hamiltonian.
  
The above assumptions can be simple written as $T_{c}=T_{c}(n,\Delta_0)$
and $\Delta_{0}=\Delta_{0}(P)$. On Fig.1 we plot two curves to study
how $T_c$ changes with  $\Delta_0$.
If we now perform an expansion of 
$T_C$ in terms of P, we obtain terms up to the third order and each
coefficient is given by

\begin{equation}
\alpha_Z=({\partial \over \partial \Delta_{0}}{\partial \Delta_0 \over
 \partial P} + {\partial \over \partial n}{ \partial n \over \partial P})^Z
 T_(n,\Delta_0)
\label{coeff}
\end{equation} 
where $Z=0,1,2$ and $3$ is the order of the corresponding 
coefficient. We can derive simple analytical expressions for
each coefficient in an approximate way,  using the universal 
parabolic fitting and with $T_{c}^{max}(P)=2\Delta_{0}(P)/\gamma$ 
which explicits the $T_c$ dependency on P (assumption ii).
This procedure gives an intrinsic term which is 
radically different than that derived on the basis of the PICT alone
as well as a new third order term. 

To illustrate the entire method we will apply it to the 
mercury system. The mercury family of compounds represents a 
real challenge to any theory for the following reasons: (a)- The
highest transition temperatures obtained up to date have been
measured on Hg1223 at 25-30 GPa\cite{Chu,MNR,Gao} reaching 
values up to 164K. (b)-The set of data for the underdoped and 
overdoped regime for the $HgBa_2CuO_{4-\delta}$ (Hg1201) could
not be interpreted\cite{Cao} by the simple PICT models. (c)- The
largest pressure effect on $T_c$ with  a change of 
almost 50K over a span of 20 GPa has been
recently  measured\cite{Acha} in the Hg2212.

We  apply now the above general procedure to the low pressure 
results of Cao et al\cite{Cao} for the Hg1201. Thus to compare
with their data
we fix the linear term at $n_{op}=0.16$, $\alpha_1=1.85$K/GPa (which 
is equal to the intrinsic term and from this we determine 
$\partial \Delta_0 / \partial P$)
and at $n=0.06$ we take $\alpha_1=2.6$K/GPa (the intrinsic term is
then $0.9$K/GPa and the charge transfer is equal $1.7$K/GPa what gives  
$\partial n/\partial P =1.8$x$10^{-3}$).
Thus the value that we obtain to $\partial n/\partial P $
is very close to other calculations 
\cite{Almasan,Gupta,Acha} and in this way
all the coefficients of  Eq.\ (\ref{coeff}) are determined.  
Far from $n_{op}$ the charge transfer term  dominates over the 
intrinsic one and the linear term 
decreases up to -1.0K/GPa  (at n=0.26) in the overdoped region. 
Our  results for all densities $n$  are in  excellent agreement with the
experimental data in both the underdoped and overdoped region, 
as shown on Figs.1a and 1b. 
\begin{figure}
\hspace*{0.5cm} \psfig{file=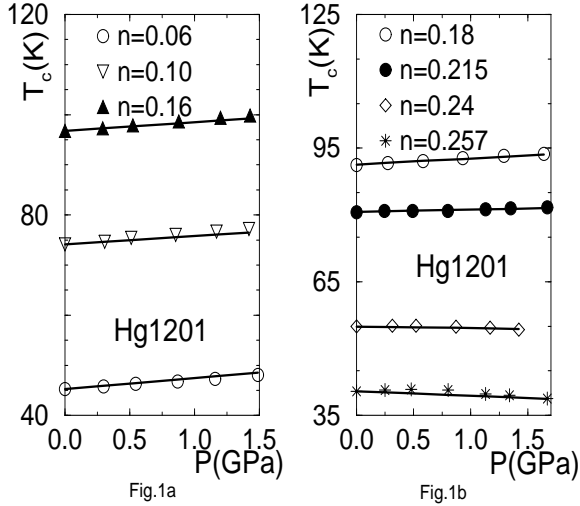,width=8cm,height=7cm}\\
\caption{a) The lines are our caculations for the underdoped 
region in comparison of the experimental points of Ref.6.
b)  The same our caculations for the overdoped 
region in comparison of the experimental points of Ref.6.}
\end{figure}
In Fig.2 we
show the results for  the high pressures measurements
for the three family of compounds of mercury\cite{Gao} at $n_{op}$. At the high
pressures the quadratic and the cubic terms in the pressure expansion become
important (for $P>20$ GPa) and  the agreement with the data
is also remarkable. It is very interesting that the values 
obtained above at the low pressures yielded results with the maxima
around 30GPa which is the value of highest $T_c$ for a HTSs\cite{Gao}.
For the first time a simple theory
is capable to describe successfully all this ensemble of data.
\begin{figure}
\hspace*{0.0cm} \psfig{file=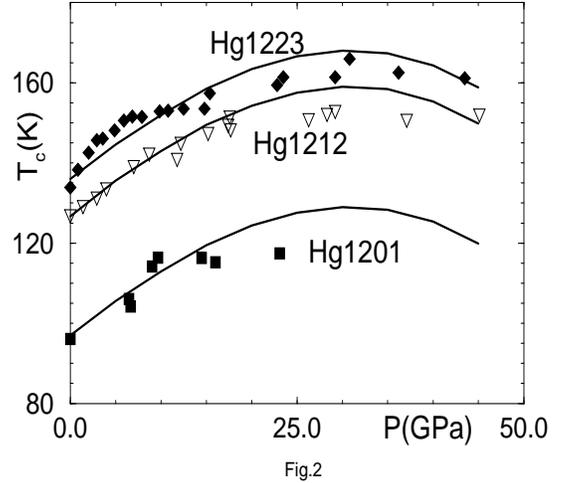,width=8cm,height=7cm}\\
\caption{  The high pressures data of Ref.11  and our
calculations (the continuous lines).}
\end{figure}

Thus, we  conclude pointing out that our novel calculations (based on
a BCS type mean field with the extended Hubbard Hamiltonian) demonstrate
to be highly successful to describe the effects of the pressure with
just two simple assumptions. The PICT which is  well accepted and 
that of the pressure induced variation  of the 
superconductor gap which was introduced, 
to our knowledge, in this work. We hope that this assumption
can be confirmed in  the future by in situ experiments like 
specific heat and tunneling measurements.
As this work  came to completion, we learnt about the  work of
Angiella et al\cite{Angiella} which also uses the extended Hubbard
model within a BCS type approach.  Furthermore they estimate the 
effects of the pressure on the attractive potential V
which goes along with the lines of our assumption over $\Delta_0$.
They also obtained very good results for the effects of pressure
in the Bi2212 family, given more support to the BCS mean field
calculations with the extended Hubbard as a model for the
the physics of the charge carriers in the  HTSs.

\end{document}